\documentclass[prb, preprint]{revtex4-1}
\usepackage{amsmath}
\usepackage{amsfonts} 
\usepackage{graphicx}
\begin{document}

\title{Modeling and experimentation with asymmetric rigid bodies: a variation
on disks and inclines}

\author{Lisandro A. Raviola}
\email{lraviola@ungs.edu.ar}
\author{Oscar Z\'arate}
\email{ozarate@ungs.edu.ar}
\author{Eduardo E. Rodr\'iguez}
\email{erodrigu@ungs.edu.ar}
\affiliation{Instituto de Industria, Universidad Nacional de General Sarmiento. 
Juan Mar\'ia Gutierrez 1150, CP B1613GSV, Los Polvorines, Buenos Aires,
Argentina.}

\date{\today}

\begin{abstract}
We study the ascending motion of a disk rolling on an incline when
its center of mass lies outside the disk axis. The problem is suitable
as laboratory project for a first course in mechanics at the undergraduate
level and goes beyond typical textbook problems about bi-dimensional
rigid body motions. We develop a theoretical model for the disk motion 
based on mechanical energy conservation and compare its
predictions with experimental data obtained by digital video recording.
Using readily available resources, a very satisfactory agreement is obtained
between the model and the experimental observations. These results
complement previous ones that have been reported in the literature
for similar systems.
\end{abstract}


\maketitle

\section{Introduction}

Within the framework of elementary university physics, a deeper understanding
of physical concepts can be stimulated by showing direct connections
between theoretical and experimental work. Besides, it could be beneficial
to pose typical problems with some modifications in order to challenge
students skills and widen the possibilities of theoretical and experimental
analysis without adding complications exceeding the course scope.
From this perspective, we develop in this paper the investigation
of a modified version of a textbook classic: a disk rolling over an
inclined plane. \cite{Resnick-Halliday} The modification consists
in attaching a mass of small dimensions to the periphery of the disk,
shifting the center of mass to a position outside the disk axis and breaking
the original cylindrical symmetry (see Fig. \ref{fig:esquema}).
The first historical reference we found about this topic is the book
of A. Good,  \cite{TomTit} a collection of popular science articles
of the late nineteenth century. The problem is presented there as
an amusing scientific curiosity because, under appropriate circumstances,
the disk ascends along the inclined plane from an initial state of
rest, which in the author's own words ``seems to contradict the immutable
laws of gravity.'' Based on this simple and easily reproducible experience,
it is pertinent to ask: 
\begin{enumerate}
\item Under what conditions will the disk rise the incline from an initial
state of rest?
\item If the disk rises, in what way does it and how far can it go? 
\end{enumerate}
These questions can motivate a useful laboratory project within the
curricular boundaries of a first mechanics course for students of
science and engineering. The problem favors a combined application
of theoretical background (mechanical energy conservation, 2D dynamics
of rigid bodies) and experimental skills (device construction, measurement,
data fitting and analysis). As an added value, the proposal can be
carried out using experimental resources usually found in teaching
laboratories, and freely available software. 

Although the analysis of the motion of asymmetric rigid bodies is
not new in the literature, \cite{Theron-AJP-2000,Theron-Maritz-MathCompModel-2008,Maritz-Theron-AJP 2012,Carnevali-May-AJP-2005,Taylor-Fehrs-AJP-2010,Gomez-EJP-2012} our treatment is complementary in both methodology and results. Previously, Carnevali and May \cite{Carnevali-May-AJP-2005} investigated a similar
problem from a Lagrangian point of view. Their approach allowed them to
obtain the temporal evolution of the relevant kinematic variables
at the cost of exceeding the possibilities of an introductory course,
a difficulty we want to avoid in the present paper. The set of works of Theron and Maritz
\cite{Theron-AJP-2000,Theron-Maritz-MathCompModel-2008,Maritz-Theron-AJP 2012}
conforms a comprehensive study of related systems, by means of vector
and energy methods. These authors developed a sophisticated model including friction effects, and
highlighted the variety of possible motions (rolling, slipping, skidding
and hopping) applying analytical and numerical techniques.\cite{Theron-AJP-2000,Theron-Maritz-MathCompModel-2008}
Lately, they experimentally confirmed that their model captures the essential
aspects of the motion of an asymmetric hoop on a horizontal plane. \cite{Maritz-Theron-AJP 2012}
In a similar way, Taylor and Fehrs  verified Theron and Maritz' theoretical predictions regarding 
the hopping conditions, \cite{Taylor-Fehrs-AJP-2010} as well as G\'omez \emph{et al}. based on an
independent model. \cite{Gomez-EJP-2012} 

In all the aforementioned papers, particular attention was devoted
the case of highly eccentric bodies $(\gamma>0.5)$ in descending
or horizontal motions, as this favors the interesting hopping phenomenon,
but the intriguing upward motion was not analyzed. While the theoretical
tools used by these authors fall within the baggage of an elementary
course, the resulting model acquires considerable complexity. By contrast,
in the present paper we apply energy methods that lead to a theoretical
model more adequate to teaching purposes. We also analyze the case
of slightly eccentric bodies $\left(\gamma<0.5\right)$ and investigate
the necessary conditions for the ascending motion. Our aim is twofold:
first, to develop a specific didactic proposal that integrates both
theoretical and experimental issues; and second, to provide original
results regarding the system under study.

To measure the kinematic variables and various system parameters,
we resort to digital video techniques,  \cite{Gil-AJP-2006,Brown-TPT-2009}
taking advantage of the ubiquitous presence of computers and digital
cameras in today's teaching laboratories. The digital recording of
a mechanical system motion provides high quality information about
its variables, while the subsequent analysis of data is facilitated
by the wide range of open source software that teachers and students
can download and use for free. In our case, we used \emph{Tracker}\cite{Tracker-OSP} 
for video analysis, and \emph{Python }\cite{Python} for the numerical 
calculations and graphics. 

In the next section we develop the theoretical model for the phenomenon
and derive some predictions that undergo experimental testing in sections
III and IV. The last section states general conclusions based on the
results obtained. 

\section{Theoretical model}

\begin{figure}[h]
\noindent 
\includegraphics[scale=0.5]{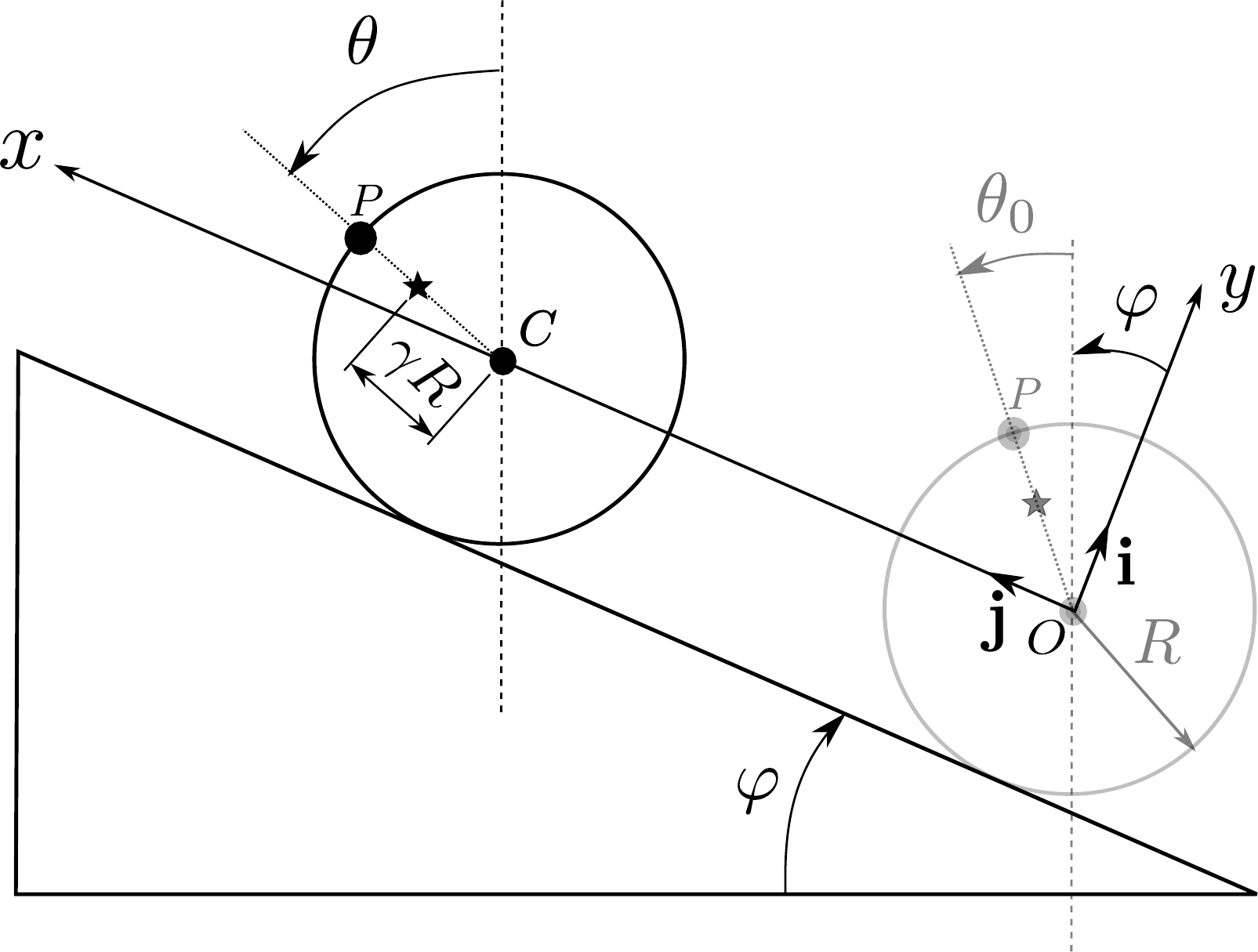}
\caption{\label{fig:esquema}Schematic diagram of the problem showing the relevant parameters and the chosen reference frame. }

\end{figure}

\noindent A disk with center $C$, radius $R$ and mass $M_{D}$ with
a particle $P$ of mass $M_{P}$ attached to its perimeter begins
to move from an initial state of rest over an inclined plane of angle
$\varphi$. At the initial time, segment $CP$ forms an angle $\theta_{0}$
with respect to the vertical direction. The reference frame origin
$O$ is located at the initial position of the disk geometric center,
and $x$ is the distance covered by $C$ in the direction parallel
to the inclined plane. The coordinate system and some parameters of
the model are shown in Fig. \ref{fig:esquema}. 

Our theoretical model stems from the \emph{rolling without slipping}
condition, given by the constraint equation
\begin{equation}
\dot{x}-R\dot{\theta}=0\label{eq:rolling_cond}
\end{equation}
whose integrated form for the initial conditions $x(0)=\dot{x}(0)=0$
is 
\begin{equation}
x=R\left(\theta-\theta_{0}\right)\label{eq:rolling_cond_int}
\end{equation}
If we neglect aerodynamic drag, the previous condition is equivalent
to mechanical energy conservation. \cite{Resnick-Halliday} This hypothesis
simplifies the analysis of the problem, sidestepping the (nonlinear)
equations of motion for $x(t)$ and $\theta(t)$, whose derivation
is generally beyond the scope of the course. Consequently, we resort
to the first integral 
\begin{equation}
E=T+V=\mathrm{const.}\label{eq:energy_cons}
\end{equation}
relating the system's mechanical energy $E$, its gravitational potential
energy $V$ and its kinetic energy $T$, in order to obtain theoretical
relations between physical quantities characterizing the phenomenon.
Predictions arising from this hypothesis will be subsequently analyzed
through experiments. 

To simplify the ensuing discussion, we define the \emph{eccentricity
parameter} $\gamma$ as 

\begin{equation}
\gamma=\frac{R_{\star}}{R}=\frac{M_{P}}{M}\label{eq:gamma}
\end{equation}
where $M=M_{P}+M_{D}$ is the total mass of the system and $R_{\star}$
is the distance from $C$ to the center of mass, indicated in figure
\ref{fig:esquema} with the symbol $\star$.

The potential energy as measured from $O$ is a function of the angular
variable $\theta$ (other symbols represent fixed parameters and initial
conditions for each particular case) 
\begin{eqnarray}
V(\theta) & = & Mg\left(x\ \sin\varphi+\gamma R\ \cos\theta\right)\nonumber \\
 & \overset{\eqref{eq:rolling_cond_int}}{=} & MgR\left(\left(\theta-\theta_{0}\right)\sin\varphi+\gamma R\ \cos\theta\right)
\end{eqnarray}
Eliminating needless constant terms, the potential energy formula
can be cast in the simpler form 
\begin{equation}
V(\theta)=MgR\left(\theta\sin\varphi+\gamma\cos\theta\right)\label{eq:potential_nrg}
\end{equation}
The body will ascend from its initial state of rest if the potential
energy decreases as a function of $\theta$, allowing an increase
of kinetic energy. Therefore, we arrive at the\emph{ }necessary condition
for ascending motion 
\begin{equation}
\frac{dV}{d\theta}(\theta_{0})=MgR\left(\sin\varphi-\gamma\ \sin\theta_{0}\right)\leq0
\end{equation}
implying that 
\begin{equation}
\sin\theta_{0}\geq\frac{\sin\varphi}{\gamma}
\end{equation}
This means that the disk will move upwards along the incline only
if the initial angle $\theta_{0}$ is greater than a minimum value
$\theta_{\mathrm{min}}$ (and smaller than $\theta_{\mathrm{max}}=\pi-\theta_{\mathrm{min}}$)
satisfying
\begin{equation}
\sin\ \theta_{\mathrm{min}}=\gamma^{-1}\ \sin\varphi\label{eq:theta_min}
\end{equation}
Equation (\ref{eq:theta_min}) is a prediction of our model which
answers question 1 and can be tested experimentally.

At the initial instant, the kinetic energy is $T(\theta_{0})=0$ and
the total mechanical energy (\ref{eq:energy_cons}) reads 
\begin{equation}
E=V(\theta_{0})=MgR\left(\theta_{0}\sin\varphi+\gamma\cos\theta_{0}\right)\label{eq:total_nrg}
\end{equation}
The total kinetic energy is the sum of the disk kinetic energy $T_{D}$
and particle's kinetic energy $T_{P}$ 
\begin{equation}
T=T_{D}+T_{P}=\left(\frac{1}{2}M_{D}\ \dot{x}^{2}+\frac{1}{2}I_{C}\ \dot{\theta}^{2}\right)+\frac{1}{2}M_{P}\ v_{P}^{2}\label{eq:kinetic_nrg}
\end{equation}
$I_{C}$ is the moment of inertia of the disk about its geometrical
center, and $v_{P}$ is the magnitude of the particle velocity given
by 
\begin{equation}
\mathbf{v}_{P}=\mathbf{v}_{C}+\boldsymbol{\omega}\times(\mathbf{r}_{P}-\mathbf{r}_{C})=\mathbf{v}_{C}+\boldsymbol{\omega}\times\boldsymbol{r}_{CP}
\end{equation}
$\mathbf{v}_{C}=\dot{x}\ \mathbf{i}=R\dot{\theta}\ \mathbf{i}$ is
the velocity of the disk center and $\mathbf{r}_{CP}=R\sin\left(\theta+\varphi\right)\mathbf{i}+R\cos\left(\theta+\varphi\right)\mathbf{j}$
is the position of $P$ relative to $C$. Since
\begin{eqnarray}
\boldsymbol{\omega}\times\mathbf{r}_{CP} & = & -\dot{\theta}\ \mathbf{k}\times R\left(\sin\left(\theta+\varphi\right)\mathbf{i}+\cos\left(\theta+\varphi\right)\mathbf{j}\right)\nonumber \\
 & = & R\dot{\theta}\left(\cos\left(\theta+\varphi\right)\mathbf{i}-\sin\left(\theta+\varphi\right)\mathbf{j}\right)
\end{eqnarray}
then
\begin{equation}
v_{P}^{2}=2R^{2}\dot{\theta}^{2}\left(1+\cos\left(\theta+\varphi\right)\right)\label{eq:p_velocity-1}
\end{equation}

Writing $I_{C}=\kappa M_{D}R^{2}$,  \cite{Kappa} taking into account
equations (\ref{eq:rolling_cond}), (\ref{eq:gamma}) and substituting
(\ref{eq:p_velocity-1}) into (\ref{eq:kinetic_nrg}) we get 
\begin{equation}
T=\frac{1}{2}M\dot{x}^{2}\left[\left(1-\gamma\right)\left(1+\kappa\right)+2\gamma\left(1+\cos\left(\theta+\varphi\right)\right)\right]\label{eq:kinetic_nrg_2}
\end{equation}
After a little algebra, equations (\ref{eq:energy_cons}), (\ref{eq:potential_nrg}),
(\ref{eq:total_nrg}) and (\ref{eq:kinetic_nrg_2}) lead to

\begin{equation}
\dot{x}=\sqrt{-2gR\frac{\left(\theta-\theta_{0}\right)\ \sin\varphi+\gamma\left(\cos\left(\theta\right)-\cos\left(\theta_{0}\right)\right)}{\left(1-\gamma\right)\left(1+\kappa\right)+2\gamma\left(1+\cos\left(\theta+\varphi\right)\right)}}\label{eq:phase_curve}
\end{equation}
Since $\theta=\frac{x}{R}+\theta_{0}$, we arrived at a relation between
$\dot{x}$ and $x$ that can be experimentally verified, answering
question 2 in \emph{phase space }$(x,\dot{x})$. It is worth notice
that equation (\ref{eq:phase_curve}) enforces a non-trivial constraint
between kinematic variables, initial conditions and parameters of
the model.

\section{Experimental setup}

In order to test the predictions of our model --equations (\ref{eq:theta_min})
and (\ref{eq:phase_curve})-- we implemented the experimental device
shown in Fig. \ref{fig:device}. The \textquotedblleft{}disk\textquotedblright{}
was built from two vynil records (known as Long Play, LP) with radius
$R=0.15\ \mathrm{m}$ connected by a bolt, nuts and washers through its center.
The union was strengthened by placing two CDs in the central area
of each LP. The total mass of the resulting structure was $M_{D}=0.303\ \mathrm{kg}$,
and the inertia parameter calculated for this geometry was $\kappa=0.45$,
slightly below the value corresponding to a homogeneous disk due to
the extra mass positioned near the center. The ``particle'' consisted
of a screw secured by nuts on the LPs periphery and perpendicular
to both. This screw was introduced to increase the rigidity of the
device and also as a holder for placing masses of different values,
allowing modifications of the parameter $\gamma$. The ramp used was
a wooden board of length $L=0.877\ \mathrm{m}$. 

The kinematic variables $x,\dot{x}$ and parameters $\varphi,\theta_{0}$
were obtained by recording the disk motion through a digital camera
with a resolution of $1280\times720$ pixels, at a rate of 30 frames
per second $(\Delta t=1/30\ \mathrm{s})$. The camera was mounted on a tripod
to ensure its stability and correct alignment.\cite{Camera} The
framing chosen resulted from a trade-off between minimizing the distortion
due to perspective (for which the camera should be as far as possible)
and maximize the size of the object, in order to reduce the uncertainty
of points positions. \cite{Gil-AJP-2006,Brown-TPT-2009} As a control procedure,
we measured two identical rules in different positions of the frame
and in mutually perpendicular directions, verifying that deformations
due to perspective were negligible. A plumb line in the center of
the scene provided the vertical reference direction for measuring
angles, and the length $L$ of the table was taken as reference for
distances. With this configuration, we recorded the motions corresponding
to different parameter sets $(\gamma,\varphi,\theta_{0})$.

In each video, we determined manually at every frame the $x$ coordinate
of the disk center using the software \emph{Tracker. }We also measured
from video the tilt angle $\varphi$ and the initial angle $\theta_{0}$
corresponding to the moment when the disk is released and begins its
motion. The estimated uncertainties for these parameters were $\Delta\varphi=0.4^{\circ}$
and $\Delta\theta_{0}=1.5^{\circ}$, respectively. Fig. \ref{fig:device}
shows the graphical user interface of the software during the video
analysis stage. It can be seen the coordinate system axes (whose orientations
agree with Fig. \ref{fig:esquema}), the paths of the disk center
and the particle (describing a cycloid), the reference length and
the vertical direction given by the plumb. The video and analysis
files can be found at the authors website. \cite{Videos}

\noindent \begin{center}
\begin{figure}[h]
\noindent \begin{centering}
\includegraphics[scale=0.36]{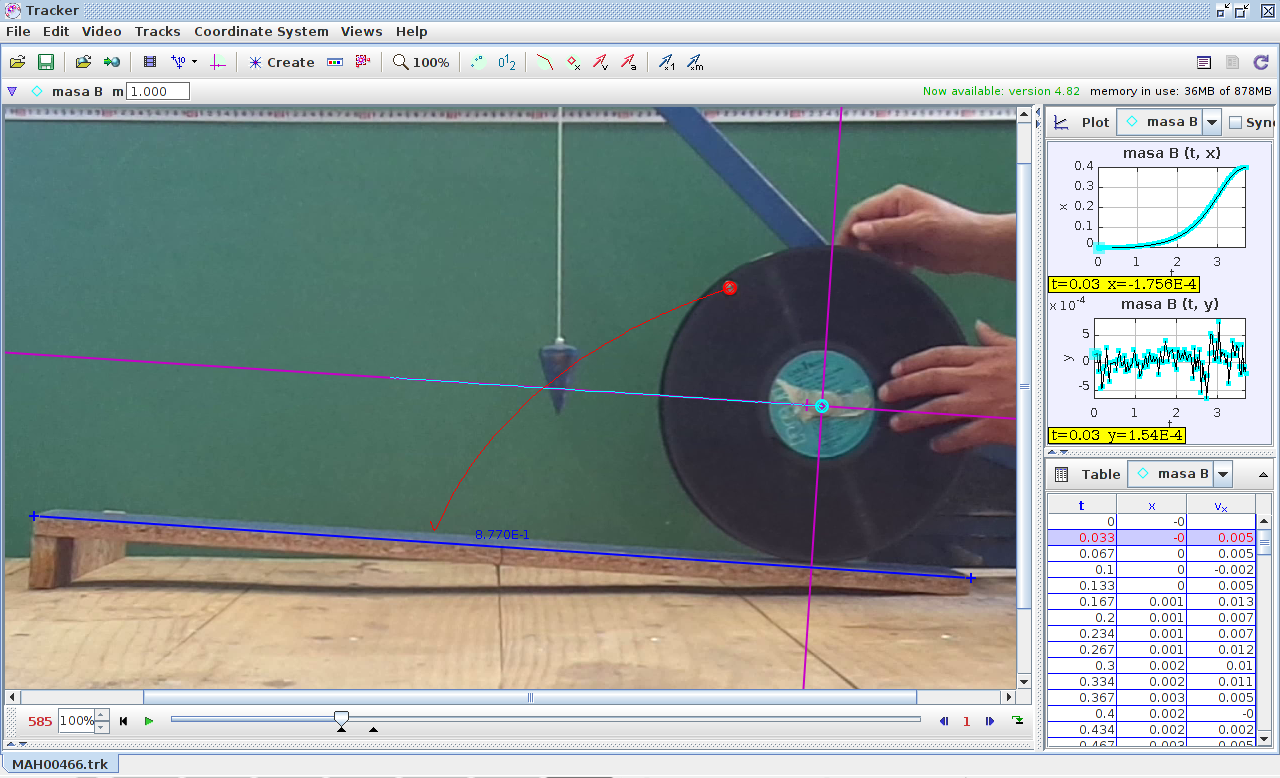}
\par\end{centering}
\caption{\label{fig:device}Experimental setup as seen inside the \emph{Tracker} graphical user interface. The ramp is used as the reference length (blue segment). The $x$ axis is parallel to the ramp (in magenta). It can be noted that the path followed by the disk center (in cyan) has the same direction as the $x$ axis. The attached particle's path (in red) is a cycloid, as expected from the rolling without slipping condition. The plumb line gives a fixed reference direction for measuring angles.} 
\end{figure}

\par\end{center}

\section{Results}

\begin{figure}[h!]
\noindent \begin{centering}
\includegraphics[scale=0.7]{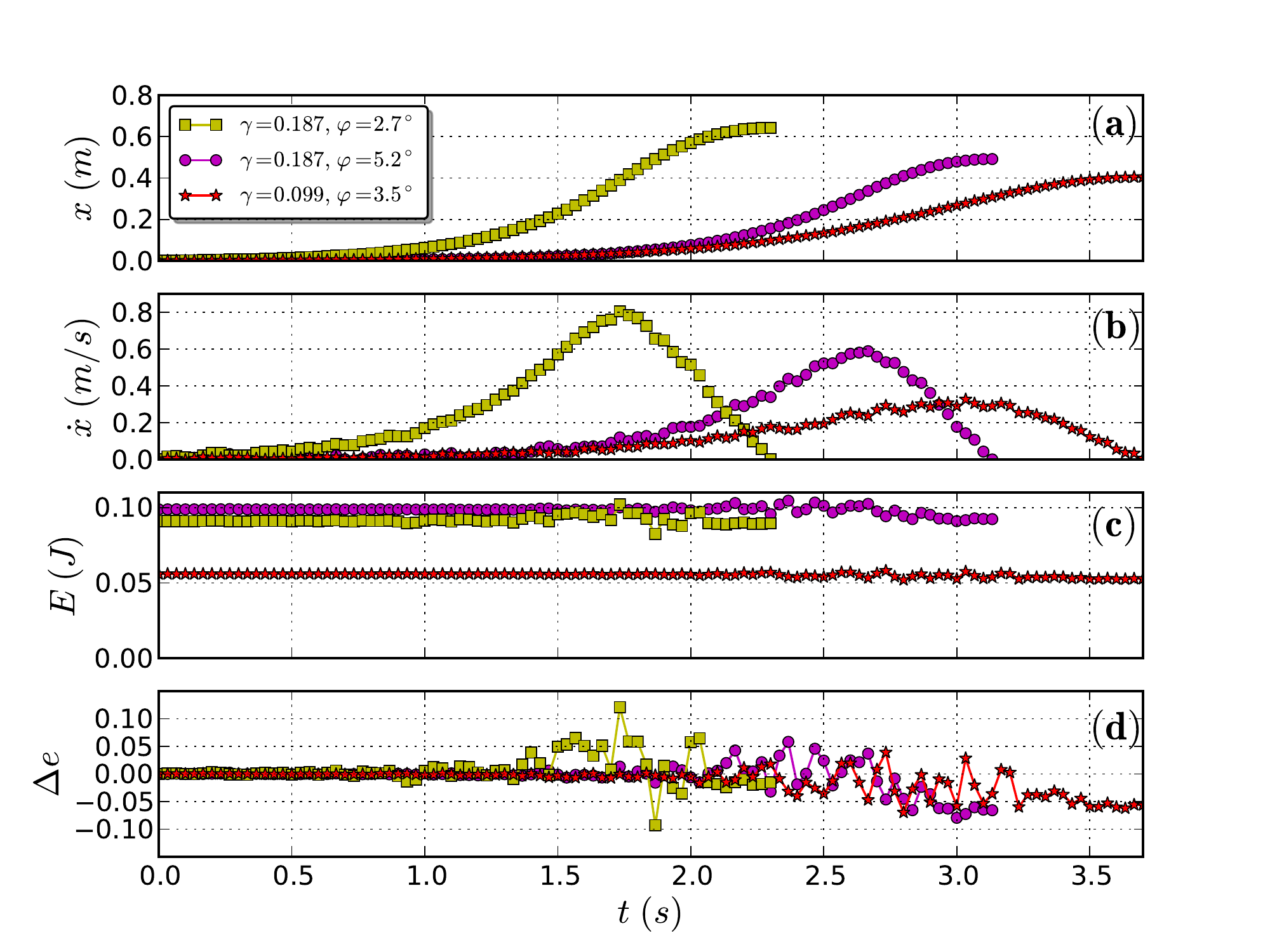}
\par\end{centering}

\noindent\caption{\label{fig:versus_t}Position, velocity and energy evolution for various representative cases. 
Panel $(a)$ shows the $x$ coordinate of the disk center as a function of time. Panel $(b)$ is the velocity of the disk center 
as calculated from data in $(a)$ by numerical differentiation. Panel $(c)$ displays the mechanical energy for the same data, and $(d)$ exhibits the time evolution of the relative energy variation with respect to the initial mechanical energy.}
\end{figure}

\noindent Fig. \ref{fig:versus_t} summarizes the information we
extracted from videos to validate our model and answer the initial
motivating questions. Panel $(a)$ of this figure shows the measured
$x$ coordinate as a function of time $t$ for various typical cases.
From this data, we calculated the speed of the disk center by means
of numerical differentiation using the midpoint rule \cite{NoteDeriv} $\dot{x}(t_{i})=\frac{1}{2\Delta t}\left(x(t_{i}+\Delta t)-x(t_{i}-\Delta t)\right)$.
The resulting values are displayed in panel $(b)$
within the same figure. With these values of ($x$,$\dot{x}$), we
checked the assumption of mechanical energy conservation upon which
the model rests, by calculating the kinetic and potential energy at
each frame. It is worth pointing out that normally this is an issue
not addressed experimentally within the subject of 2D rigid body motions
in basic mechanics courses, and is only treated theoretically from
the viewpoint of Coulomb's classic friction model. With this purpose,
in panel $(c)$ we exhibit the temporal evolution of mechanical energy
for the previous cases. When comparing initial and final absolute
values we can see a slight decrease of mechanical energy. For a better
appreciation, this change is assessed by calculating the relative
variation $\Delta e(t_{i})=\left(E(t_{i})-E(0)\right)/E(0)$, whose
evolution is shown in panel $(d)$. These variations are on average
of about 5\% or less, as reported by Carnevali 
for low speeds, \cite{Carnevali-May-AJP-2005} and only exceptionally reach 10\%. The reason for
the extreme values of the relative variation is the following: at
high speeds the image of the disk center becomes ``fuzzy'' due to
the finite time of image capture between frames, increasing the uncertainty
in position and speed, and consequently in energy. This explains the
wild oscillations and the local deviation from the average behavior
near maxima of $\dot{x}$ which can be seen in $(d)$. In view of
these results we conclude that the hypothesis of conservation of mechanical
energy (\ref{eq:energy_cons}) holds reasonably well for our purposes.

\begin{figure}[h!]
\noindent \begin{centering}
\includegraphics[scale=0.7]{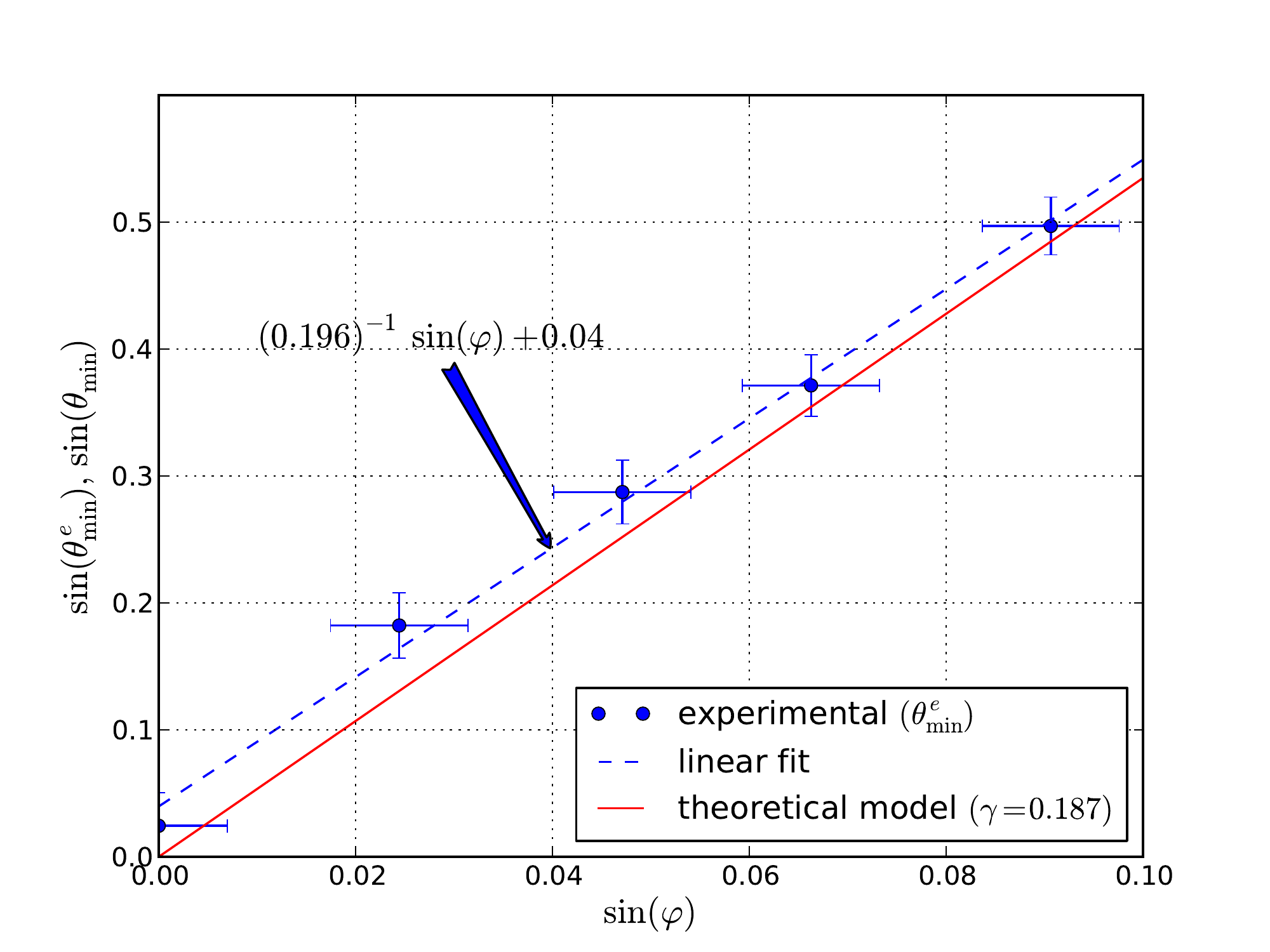}
\end{centering}
\caption{\label{fig:theta_min} Sine of the theoretical ($\theta_\mathrm{min}$) and experimental ($\theta_\mathrm{min}^e$) values of the minimum initial angle as a function of the sine of plane inclination $\varphi$ for the observed value of $\gamma = 0.187$. The solid (red) line is the theoretical model prediction for this value of $\gamma$, given by equation (\ref{eq:theta_min}). (Blue) circles indicate measured data. The discontinuous (blue) line represents the linear fit of the experimental values, whose equation is shown.}
\end{figure}

To verify the prediction (\ref{eq:theta_min}), we experimentally
determined the minimum initial angle from the vertical $\theta_{\mathrm{min}}^{e}$
at which the disk begins to move upwards, for various values of the
inclination angle $\varphi$. In Fig. \ref{fig:theta_min}, the pairs
$\left[\sin(\varphi),\sin(\theta_{\mathrm{min}}^{e})\right]$ are
plotted for $\gamma=0.187$ (blue circles) and compared with the theoretical
prediction (solid red line). The linear fit (discontinuous blue line)
yields $\gamma=0.196$, which falls within the experimental uncertainty
range $\Delta\gamma=0.01$ around the value calculated using (\ref{eq:gamma}),
and a discrepancy in the y-intercept compatible with the estimated
uncertainties for the angles $(\Delta\varphi=0.4,\ \Delta\theta_{\mathrm{min}}^{e}=1.5^{\circ})$.
Consequently, we can state that the prediction is confirmed within
the error margins of the measurement, and that the answer for question
1 is given by equation (\ref{eq:theta_min}). Notice that the measured
angle $\theta_{\mathrm{min}}^{e}$ is always bigger than the theoretical
one for a given $\varphi$. This is a consequence of the manual method
used to initiate the motion in the desired (ascending) sense, considering
that the minimum angle corresponds to an unstable equilibrium.

\noindent 

\begin{figure}[]
\noindent \includegraphics[scale=0.60]{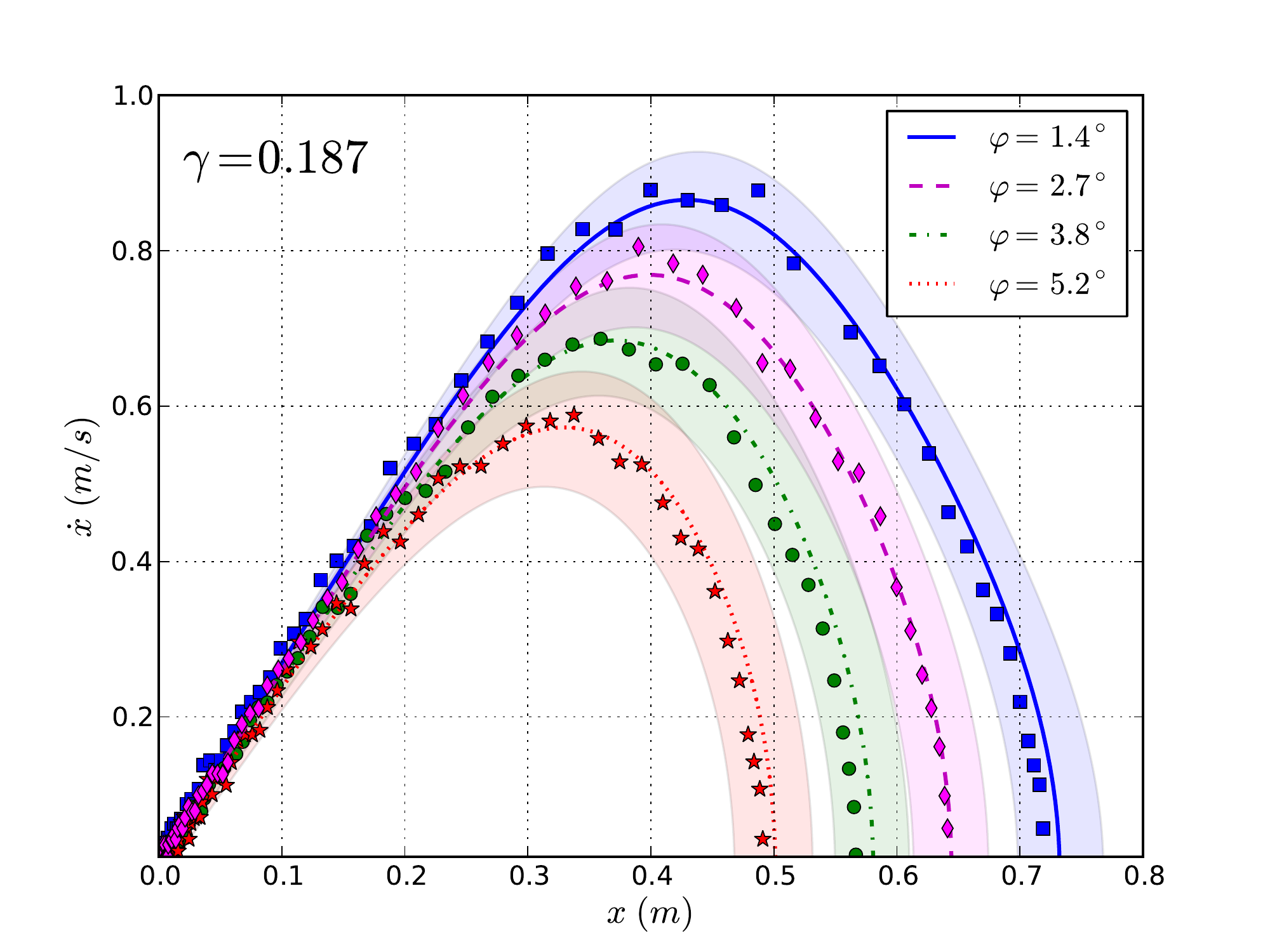}
\includegraphics[scale=0.60]{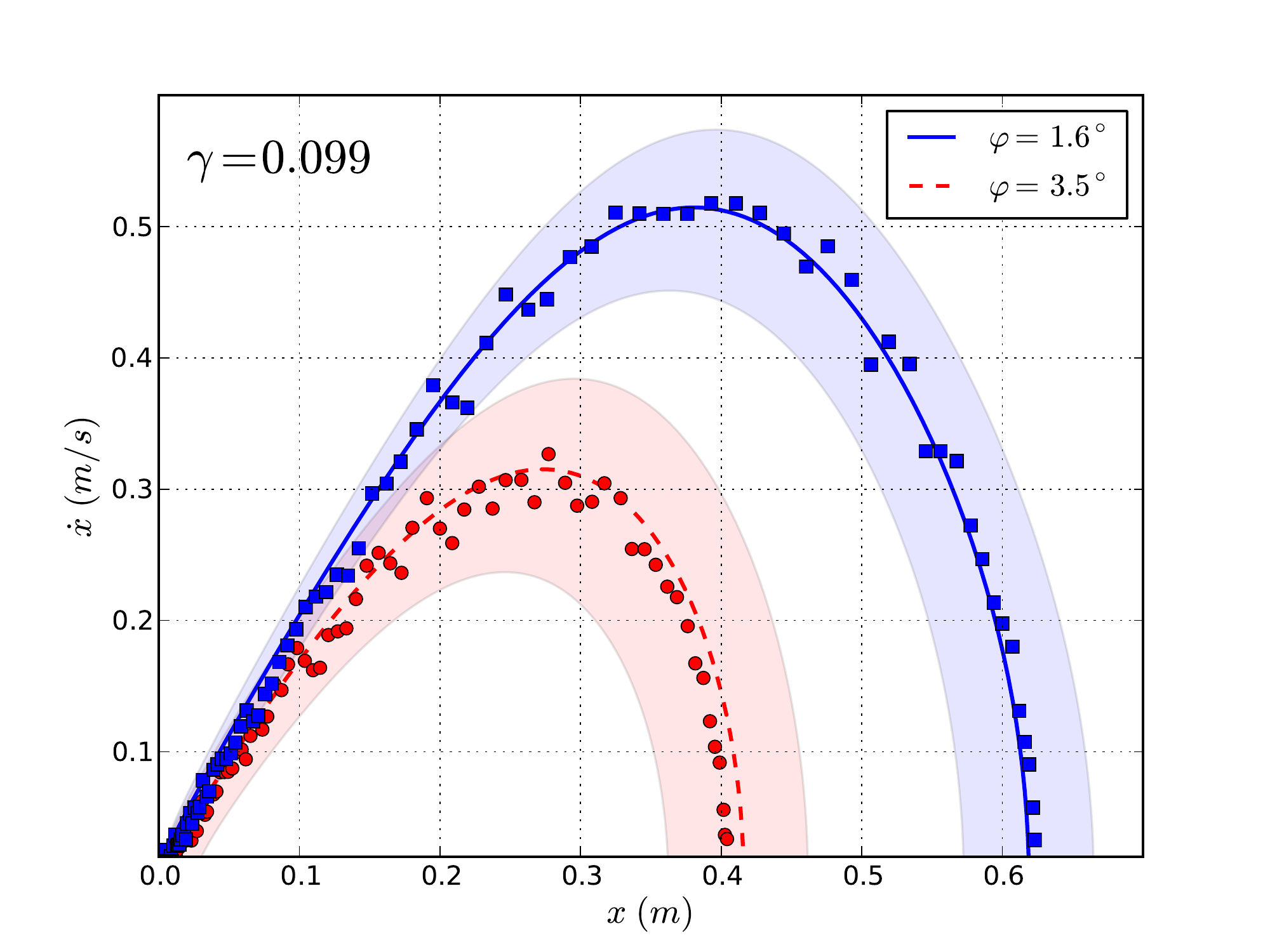}
\noindent\caption{\label{fig:phase_space}Motion of the system in phase space for different values of $\varphi$ and $\gamma$. Symbols represent the experimental values obtained from video. Lines indicate the theoretical model prediction --given by equation (\ref{eq:phase_curve})-- for the corresponding values of $\varphi$ and $\gamma$. Bands around these lines show the theoretical model uncertainty arising from uncertainties in the model parameters.}
\end{figure}

Finally, to corroborate (\ref{eq:phase_curve}) we plot in Fig. \ref{fig:phase_space} the measured pairs $(x,\dot{x})$ for several
values of the inclination angle $\varphi$ and two values of the eccentricity
parameter $(\gamma=0.094,\ \gamma=0.187)$. Lines represent
the model prediction (\ref{eq:phase_curve}) for the measured values
of $\gamma,\ \theta_{0}$ and $\varphi$. The color bands centered
at these lines indicate the theoretical prediction uncertainty arising
from uncertainties in the measurement of parameters. In all the cases
investigated, the model explains well the experimental values, as can be
observed in both figures. We conclude that equation (\ref{eq:phase_curve})
answers question 2 in phase space.

\noindent 

\section{Conclusions}
In this paper, we presented a modification of a textbook problem in
rigid body dynamics which can be used to integrate theoretical and
experimental issues in a first university mechanics course. The proposal
may be implemented as a short laboratory project, motivated by an
easily reproducible demonstration of the phenomenon and the simple
questions posed in the introduction. Besides, we obtained very satisfactory
original results resorting only to readily available tools and methods.
We confirmed the model predictions with an accuracy compatible with
the quality of the experimental observations, complementing previous
results reported in the literature for similar systems. The same methodology
can be applied to other variants of traditional problems in mechanics,
being a valuable resource in elementary physics teaching.

\section*{Acknowledgements}

The authors thank the staff of the Engineering Laboratory of the Universidad
Nacional de General Sarmiento for providing materials and physical
space to perform the experiments. This work was done under project
UNGS-IDEI 30/4045 ``Experimentos en contexto para la ense\~nanza y
el aprendizaje de la ciencia y la tecnolog\'ia.''


\begin{thebibliography}{10}
\bibitem[14]{Camera}For the same purpose could be used any camera,
phone or tablet properly aligned and with fixed focus, as auto-focus
can lead to undesirable changes between frames.
\end{thebibliography}
\end{document}